\documentclass[preprint,12pt]{elsarticle}

\usepackage{amssymb}
\usepackage{graphicx}
\usepackage{subfigure}

\journal{Journal of Sound and Vibration}

\begin{document}

\begin{frontmatter}



\title{Effective mass overshoot in single degree of freedom mechanical systems with a particle damper}


\author[1,2]{Mart\'{\i}n S\'anchez}
\author[1]{Luis A. Pugnaloni\corref{cor}}\ead{luis@iflysib.unlp.edu.ar}
\cortext[cor]{Corresponding author. Tel.: +54-221-423-3283; fax: +54-221-425-7317}

\address[1]{Instituto de F\'{\i}sica de L\'{\i}quidos y Sistemas Biol\'ogicos (CONICET La Plata, UNLP), Calle 59 Nro 789, 1900 La Plata, Argentina.}
\address[2]{Departamento de Ingenier\'{\i}a Mec\'anica, Facultad Regional La Plata, Universidad Tecnol\'ogica Nacional, 60 esq. 124 S/N, 1900 La Plata, Argentina.}

\begin{abstract}
We study the response of a single degree of freedom mechanical system composed of a primary mass, $M$, a linear spring, a viscous damper and a particle damper. The particle damper consists in a prismatic enclosure of variable height that contains spherical grains (total mass $m_\mathrm{p}$). Contrary to what it has been discussed in previous experimental and simulation studies, we show that, for small containers, the system does not approach the fully detuned mass limit in a monotonous way. Rather, the system increases its effective mass up and above $M+m_\mathrm{p}$ before reaching this expected limiting value (which is associated with the immobilization of the particles due to a very restrictive container). Moreover, we show that a similar effect appears in the tall container limit where the system reaches effective masses below the expected asymptotic value $M$. We present a discussion on the origin of these overshoot responses and the consequences for industrial applications.
\end{abstract}


\end{frontmatter}


\section{Introduction}
\label{intro}

Most mechanical systems, such as rotating machinery and aeronautic or aerospace structures, achieve damping through viscoelastic materials or viscous fluids. In general, viscoelastic materials and viscous fluids are very effective at moderate temperatures (less than 250$^{\circ}$C), but the performance of these is poor at low and high temperatures. Moreover, these materials degrade over time and lose their effectiveness.

In recent years, particle dampers (PD) have been studied extensively for use in harsh environments where other types of damping are not efficient. A PD is an element that increases the structural damping by inserting dissipative particles in a box attached to the primary system or by embedding grains within holes in a vibrating structure \cite{Panossian}. The grains absorb the kinetic energy of the primary system and convert it into heat through inelastic collisions and friction between the particles and between the particles and the walls of the box or hole. This results in a highly nonlinear mechanical system. PD are effective over a wide frequency range \cite{Panossian}. Moreover, PD are durable, inexpensive, easy to maintain and have great potential for vibration and noise suppression for many applications (see e.g. \cite{Simonian} and \cite{XU}).

Parameters such as the size and shape of the particles, density, coefficient of restitution, size and shape of the enclosure, and the type of excitation of the primary system, among many other features, are important in damping performance \cite{Marhadi}. Thus, appropriate treatment of the PD in a given structure requires careful analysis and design.

PD use a large number of small grains; therefore, its behavior is directly related to cooperative movements of those inside the cavity. The theoretical models derived from single particle systems \cite{Duncan} are not applicable to predict the performance of multi-particle systems. For more than 15 years, particle dynamics simulations have been used as a powerful tool for investigating the behavior of these types of granular systems \cite{Mao,Saeki,Bai,Fang}.

In previous works, particle dampers composed of containers of various sizes have been considered. In all of these works, the resonant frequency of the Single-Degree-of-Freedom (SDoF) system falls with respect to the undamped system. This is generally attributed to the addition of the mass of the particles, $m_\mathrm{p}$. At very low excitation amplitudes, the system behaves as if the entire mass of the particles was attached to the primary mass, $M$, of the container (i.e. $M+m_\mathrm{p}$). If the excitation level is increased, the resonant frequency gradually increases (and the damping performance increases). Eventually, the resonant frequency tends to the resonant frequency of the undamped system. This overall behavior has been discussed in various papers \cite{Fang,Liu}.

Yang \cite{Yang} has studied, experimentally, particle dampers under different conditions of excitation, different frequencies and variable gap size. The gap size is the free space left between the granular bed and the enclosure ceiling when the system is at rest. He has found that, under some conditions, the system may display effective masses above $M+m_\mathrm{p}$ or below $M$. However, a careful analysis of this phenomenon has not been carried out yet.

In this paper, we discuss results, obtained through simulations via a Discrete Elements Method (DEM), on the resonant frequency shift of SDoF mechanical systems with granular damping. We show that, contrary to what has been discussed previously, for small containers, the system does not approach the fully detuned mass limit in a monotonous way. Rather, the system increases its effective mass up and above $M+m_\mathrm{p}$ before reaching the expected limiting value. Moreover, there is a similar effect in the tall enclosure limit, where the system reaches effective masses below the expected asymptotic value $M$.

\section{Discrete Elements Method}
\label{DEM}

In order to simulate the motion of the particles in the enclosure of a PD we use a DEM. This scheme, first used by Cundall and Strack \cite{Cundall}, is widely used for numerical simulations of granular media \cite{Poschel}. We implement a velocity Verlet algorithm \cite{Allen} to update the positions (orientations) and velocities of the particles. Orientations are represented through quaternions to prevent numerical singularities \cite{Goldstein}.

We consider spherical soft particles. If $R_{i}$ and $R_{j}$ are the radii of two colliding particles, $\alpha = R_{i} + R_{j} - d_{ij}$ is the normal displacement or virtual overlap between the spheres, where $d_{ij}$ is the distance between the centers of the two spheres. Under these conditions, the interaction force $F_\mathrm{n}$ in the normal direction is based on the Hertz--Kuwabara--Kono model \cite{Schafer,Kruggel1}.
\begin{equation}
F_\mathrm{n} = -k_\mathrm{n}\alpha^{3/2}-\gamma_\mathrm{n}\upsilon_\mathrm{n}\sqrt{\alpha}
\label{normal}
\end{equation}
where $k_\mathrm{n}=\frac{2}{3}E\sqrt{\frac{R}{2}}(1-\upsilon^{2})^{-1}$ is the normal stiffness (with $E$ the Young's modulus, $\upsilon$ the Poisson's ratio and $R^{-1}=R_i^{-1}+R_j^{-1}$), $\gamma_\mathrm{n}$ the normal damping coefficient of the viscoelastic contact, and $\upsilon_\mathrm{n}$ the relative normal velocity.

On the other hand, the tangential force $F_\mathrm{s}$ is based on Coulomb's law of friction \cite{Schafer,Kruggel2}. We used a simplified model in which the friction force takes the minimum value between the shear damping force and the dynamic friction.
\begin{equation}
F_\mathrm{s} = -\min\left(\left|\gamma_\mathrm{s}\upsilon_\mathrm{s}\sqrt{\alpha}\right|,\left|\mu_\mathrm{d}F_\mathrm{n}\right|\right)\rm{sgn}\left(\upsilon_\mathrm{s}\right)
\label{tangential}
\end{equation}
where $\gamma_\mathrm{s}$ is the shear damping coefficient, $\upsilon_\mathrm{s}$ the relative tangential velocity between the two spheres in contact and $\mu_\mathrm{d}$ the dynamic friction coefficient. The sign function indicates that the friction force always opposes the direction of the relative tangential velocity.

The particles are enclosed in a prismatic container (the box or enclosure) built up of six flat walls with the same material properties as the particles defined through the parameters in Eqs. (\ref{normal}) and (\ref{tangential}).

\section{The SDoF model}
\label{simul}
\begin{figure}[htp]
\begin{center}
\includegraphics[width=12cm]{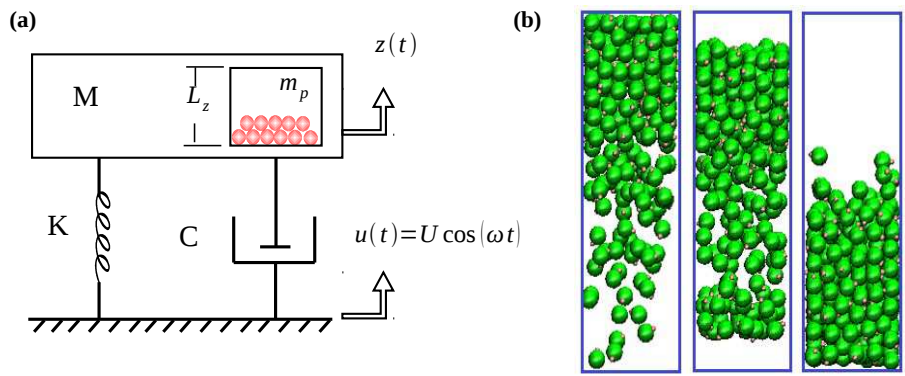}
\end{center}
\caption{\textbf{(a)} Model of the SDoF system with a particle damper. \textbf{(b)} Snapshots of the particles in the enclosure during a typical simulation.}
\label{fg:Fig. 1}
\end{figure}

Figure~\ref{fg:Fig. 1} shows the model of our SDoF system with PD which is assumed to move only along the direction of the vertical $z$-axis. The primary system consists of a mass $M = 2.37$ kg, a spring $K = 21500$ Nm$^{-1}$ and a viscous damper with damping constant $C$. We have used two different values for viscous damping, $C = 7.6$ and $26.3$ Nsm$^{-1}$. The undamped ($C=0$ and no PD) natural frequency of the primary system is $f_{0} = 15.16$ Hz.

The PD is modeled as $N = 250$ spherical grains in a prismatic enclosure of lateral side $L_x=L_y=0.03675$ m and different heights $L_z$. The material properties of the particles (and walls) and the simulation parameters are listed in Table~\ref{tab:Tabla1}. The gravitational field $g = 9.8$ ms$^{-2}$ is considered in the negative vertical direction. Although the SDoF system can only move in the vertical direction, the particles move freely inside the enclosure. 

The system is excited by the harmonic displacement of the base to which the spring and viscous damper are attached (see Fig.~\ref{fg:Fig. 1}). Let $u(t)$ and $z(t)$ be the displacement of the base and the primary mass, respectively. Then, the equation of motion for the system is given by
\begin{equation}
M\ddot{z}(t) + C \dot{z}(t) + K {z}(t) = C \dot{u}(t) + K u(t) + F_\mathrm{part}(t), \label{ec2} \ \ \ \ \ \  u(t) = U\cos(\omega t),
\end{equation}
where $F_\mathrm{part}(t)$ is the $z$-component of the force resulting from all the interactions (normal and tangential) of the enclosure walls with the particles. The amplitude, $U$, and the angular frequency, $\omega$, of the harmonic vibrating base, are control parameters.
\begin{table}[htb]
\centering
\begin{tabular}{|c|c|}
\hline Property & Value \\
\hline
\hline
Young's modulus $E$& $2.03\times10^{11}$ Nm$^{-2}$ \\
\hline
Density & 8030 kgm$^{-3}$ \\
\hline
Poisson's ratio $\upsilon$ & 0.28 \\
\hline
Friction coefficient $\mu_\mathrm{d}$ & 0.3 \\
\hline
Normal damping coefficient $\gamma_\mathrm{n}$ & $3.660\times10^{3}$ kgs$^{-1}$m$^{-1/2}$ \\
\hline
Shear damping coefficient $\gamma_\mathrm{s}$ & $1.098\times10^{4}$ kg$s^{-1}$m$^{-1/2}$ \\
\hline
Excitation amplitude $U$ & 0.0045 m \\
\hline
Time step $\delta t$ & $8.75\times10^{-8}$ s \\
\hline
Time of simulation & 13.12 s \\
\hline
Particle radius & 0.003 m \\
\hline
Total particle mass $m_\mathrm{p}$ & 0.227 kg \\ 
\hline
\end{tabular}
\caption{Material properties of the particles and simulation parameters.}
\label{tab:Tabla1}
\end{table}

We have obtained the frequency response function (FRF) for different enclosure heights $L_z$. The initial condition for each simulation consists of a simple deposition of the particles inside the enclosure, starting from a dilute random arrangement, before applying the base excitation.

\section{Data analysis}
\label{anal}

As shown in Table~\ref{tab:Tabla1}, we have simulated the vibration of the system for $13.12$ s. After an initial transient, the system reaches a steady state. This steady state can display either regular or chaotic behavior, depending on the excitation frequency, box height, etc. In all cases, the final 10\% of the time of the simulations has been used for the data analysis, which has proved to be sufficient to ensure that the steady state has been reached. We have studied frequency excitations in the range ($0.5 - 30.0$ Hz).

We carry out a simple evaluation of the \textit{effective mass} and \textit{effective damping} of the PD by fitting the FRF to a SDoF system with no particles in the enclosure. Other approximate methods such as the \textit{power flow method} used by Yang \cite{Yang} present numerical instabilities and can yield negative effective masses (see e.g. \cite{Wong}).

\begin{figure}
\begin{center}
\includegraphics[width=10cm]{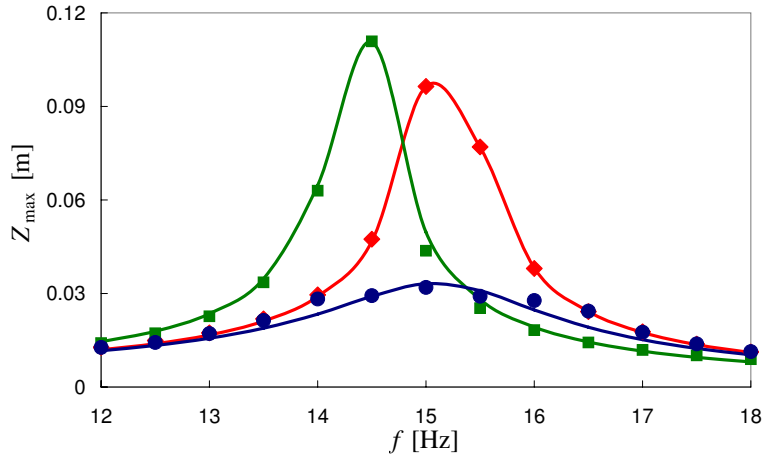}
\end{center}
\caption{Examples of the FRF of the SDoF system with PD for different $L_z$. Green squares: $L_z=0.057$ m, blue circles: $L_z=0.1225$ m and red diamonds: $L_z=0.372$ m. These results correspond to the simulations with $C=7.6$ Nsm$^{-1}$. The solid lines are fits of the FRF with an equivalent mass--spring--dashpot model without PD [see Eq. (\ref{ecsol2})].}
\label{fig:fit}
\end{figure}

The amplitude of the response $X$ of a system with no PD is given by
\begin{equation}
X = U\left[\frac{K^2+(C_\mathrm{eff}\omega)^2}{(K-M_\mathrm{eff}\omega^2)^2+(C_\mathrm{eff}\omega)^2}\right]^{1/2}
\label{ecsol2}
\end{equation}                                                                                                                                                                      
We carry out a least-squares curve fitting of the DEM data with Eq.~(\ref{ecsol2}). The values of $K$ and $U$ are fixed to the corresponding values in our simulations and $C_\mathrm{eff}$ and $M_\mathrm{eff}$ are fitting parameters. 

It is important to mention that for certain enclosure heights, the FRF is not a smooth function as the one corresponding to Eq. (\ref{ecsol2}). However, the overall shape is well described by this fit. For small and large enclosures, the shape of the FRF of the PD is rather smooth and the fits are warranted. Examples of the quality of the fits are shown in Fig. \ref{fig:fit} for three different $L_z$. An improvement of the fits for the FRF at intermediate $L_z$ for which some fluctuations are present could be achieved by using models with extra degrees of freedoms (e.g. a tuned mass damper). However, the characteristic frequencies of the PD are not well separated and the equivalent mass--spring--dashpot model constitute a good first order approximation.

\section{Results}
\label{res}

\begin{figure}
\begin{center}
\includegraphics[width=12cm]{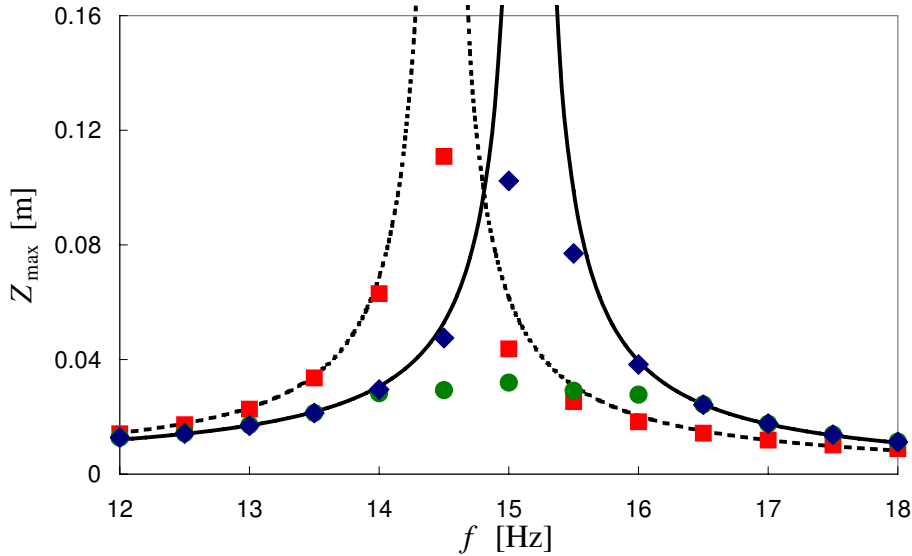}
\end{center}
\caption{Frequency response function for the SDoF system with PD. Each curve corresponds to a different box height $L_z$. Red squares: $L_z=0.057$ m, green circles: $L_z=0.1225$ m and blue diamonds: $L_z=0.372$ m. The black continuous line corresponds to the SDoF system without particles in the enclosure. The black dashed line corresponds to an equivalent system with an added mass $m_\mathrm{p}$ equal to the mass of the particles. These results correspond to the simulations with $C=7.6$ Nsm$^{-1}$. The green circles correspond to value of $L_z$ for which the maximum effective damping is obtained.}
\label{fig:frf}
\end{figure}

Examples of the FRF of the PD can been seen in Fig. \ref{fig:frf} for a few box sizes $L_z$. It is important to note that, in general, the gap size and not $L_z$ controls the dynamics. In this work all simulations are carried out with the same number of particles and particle size. Hence, the gap size is directly obtained as $L_z - h$; with $h=0.039$ m the approximate height of the granular bed at rest. For a study on the dependence of the dynamics on the number of particles and particle size see Ref. \cite{Saeki,Sanchez}. 

The general trends observed in Fig. \ref{fig:frf} are consistent with previous experimental and simulation works (see e.g. \cite{Saeki}, \cite{Fang} and \cite{Liu}). For small gaps ($L_z < 0.087$ m) the FRF tends to the response of a system without particles but with an added mass equivalent to the mass $m_\mathrm{p}$ of the particles. For very tall boxes ($L_z > 0.222$ m), the response follows the one expected for an empty container. For intermediate values of $L_z$, significant granular damping is obtained with damped resonant frequencies intermediate between the two asymptotic cases. However, these intermediate values of $L_z$ yield more complex FRFs with the presence of more than one peak \cite{Liu}.

\subsection{Effective damping and effective mass}

\begin{figure}
\begin{center}
\includegraphics[width=12cm]{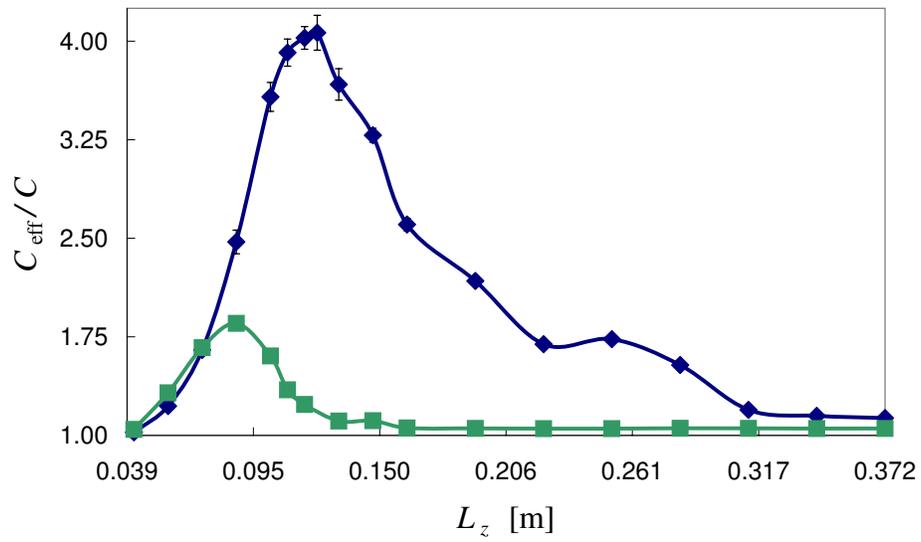}
\end{center}
\caption{Effective damping $C_\mathrm{eff}$ relative to the viscous damping $C$ as a function of $L_z$. The blue line corresponds to the SDoF system with $C=7.6$ Nsm$^{-1}$ and the green line to the system with $C= 26.365$ Nsm$^{-1}$. The error bars correspond to the asymptotic standard error of the $C_\mathrm{eff}$ best fit.}
\label{fig:ceff}
\end{figure}

A simple evaluation of the effective damping yield by the particles can be done by fitting the FRF to a SDoF system including only a viscous damper as discussed in Section \ref{anal}. The effective damping, $C_\mathrm{eff}$ is plotted as a function of $L_z$ in Fig. \ref{fig:ceff}. A clear optimum value of $L_z$ is predicted as in several previous works \cite{Saeki,Sanchez}. Notice that an increase of the viscous damping $C$ leads to a less important influence of the PD on the effective damping $C_\mathrm{eff}$. This is due to the fact that the damping due to the dashpot of the SDoF system reduces the transfer of energy to the particles inside the enclosure.
 
The two limiting cases (small $L_z$ and large $L_z$) have been explained \cite{Fang, Yang, Brennan} in terms of the proportion of the time that particles spend in contact with the enclosure. For small $L_z$, particles are essentially fixed in their positions since the constraint imposed by the walls impede their relative motion. Therefore, the particles behave as a simple mass added to the primary mass $M$ of the system and provide no extra damping (in this limit the effective damping equals the viscous damping). Conversely, if the enclosure leaves sufficient room for the motion of the particles, the granular bed will reach a gas-like state (at the large excitation levels achieved near resonance) in which most of the time particles are in the air and only occasionally collide against the floor and ceiling of the box. In this case, the particles will barely influence the motion of the primary system and $C_\mathrm{eff}$ falls back to the baseline, $C$, imposed by the viscous damper.

\begin{figure}
\begin{center}
\includegraphics[width=12cm]{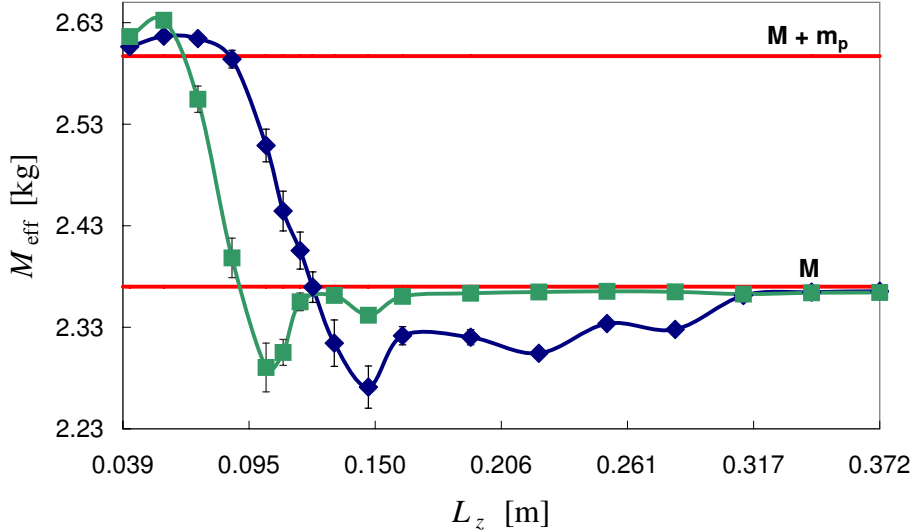}
\end{center}
\caption{Effective mass as a function of $L_z$. The blue line corresponds to the SDoF system with $C=7.6$ Nsm$^{-1}$ and the green line to the system with $C= 26.365$ Nsm$^{-1}$. The error bars correspond to the asymptotic standard error of the $M_\mathrm{eff}$ best fit.}
\label{fig:meff}
\end{figure}

In Fig. \ref{fig:meff} we plot the effective mass, $M_\mathrm{eff}$, obtained from the fits. The expected limiting masses $M$ and $M+m_\mathrm{p}$ are shown as horizontal lines for reference. Our data surveys a larger number of box heights in comparison with previous studies. We can see that, in contrast with the suggestion of previous studies, the two limiting cases are not approached in a monotonous way. The system reaches effective masses above $M+m_\mathrm{p}$ as $L_z$ is decreased and finally falls towards the limit value. Similarly, for large $L_z$, an increase of the box height leads to effective masses below $M$ before the limit value is approached. It is worth mentioning that a similar behavior has been observed in some experiments \cite{Yang}. The author reports that the resonant frequency (which is simply related to our effective mass by $M_\mathrm{eff}=k/(2\pi f_0)^2$) can reach values above the expected for an empty enclosure.

The presence of these overshoots when approaching the two limiting cases (zero gap size and infinite gap size) has been overlooked or received little attention in the past and is somewhat against intuition. In general, studies of PD are focused on the region of gap sizes where the maximum effective damping is observed. There, the effective mass is, as expected, intermediate between $M$ and $M+m_\mathrm{p}$. However, design constraints may require a PD to work at gap sizes off the optimum damping and in the region of effective mass overshoot. In what follows, we discuss the origin of such behavior in the response of the PD by considering the internal motion of the granular bed.

\subsection{Internal motion of the granular bed}

\begin{figure}[htp]
\begin{center}
\includegraphics[width=6cm]{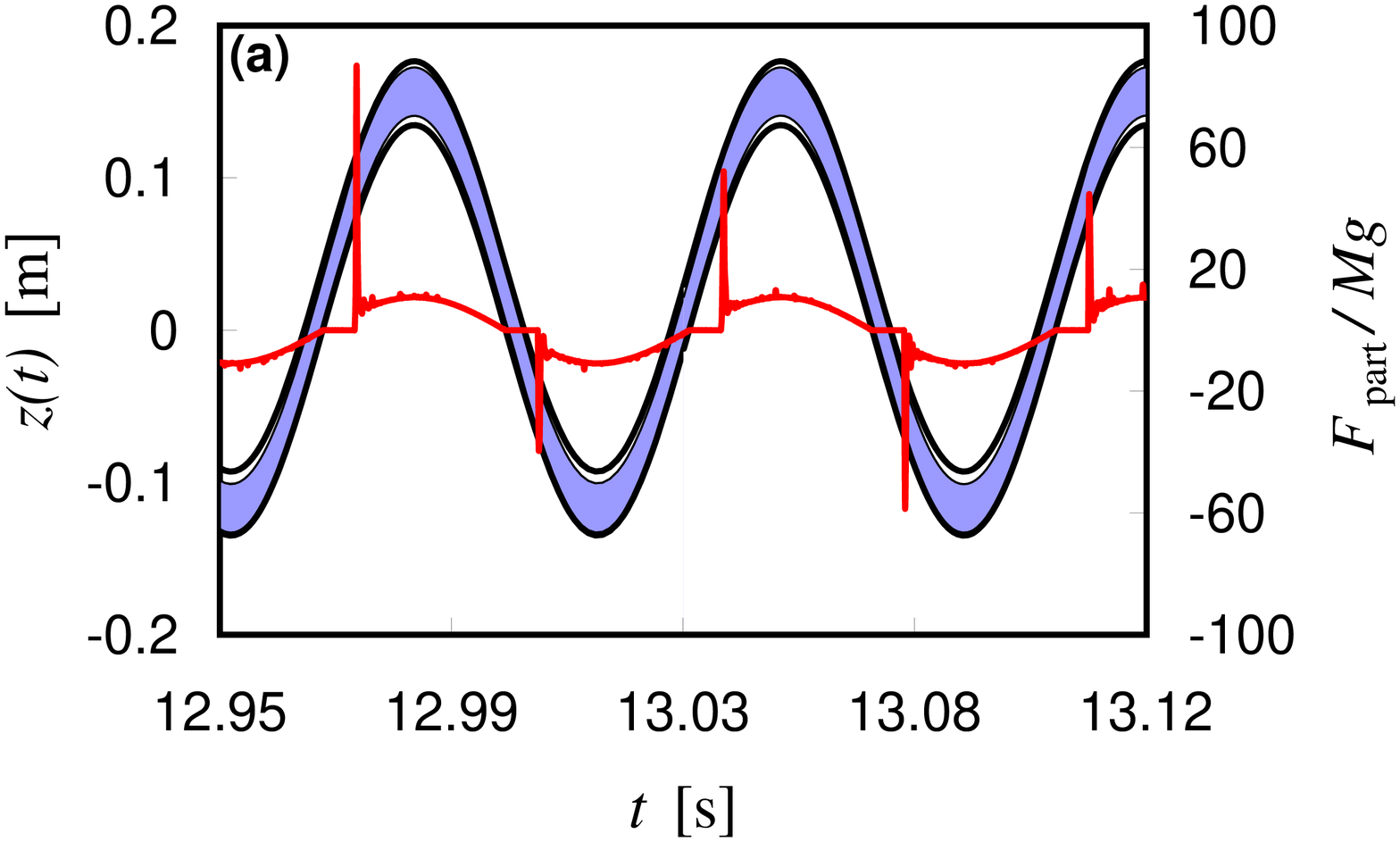}
\includegraphics[width=6cm]{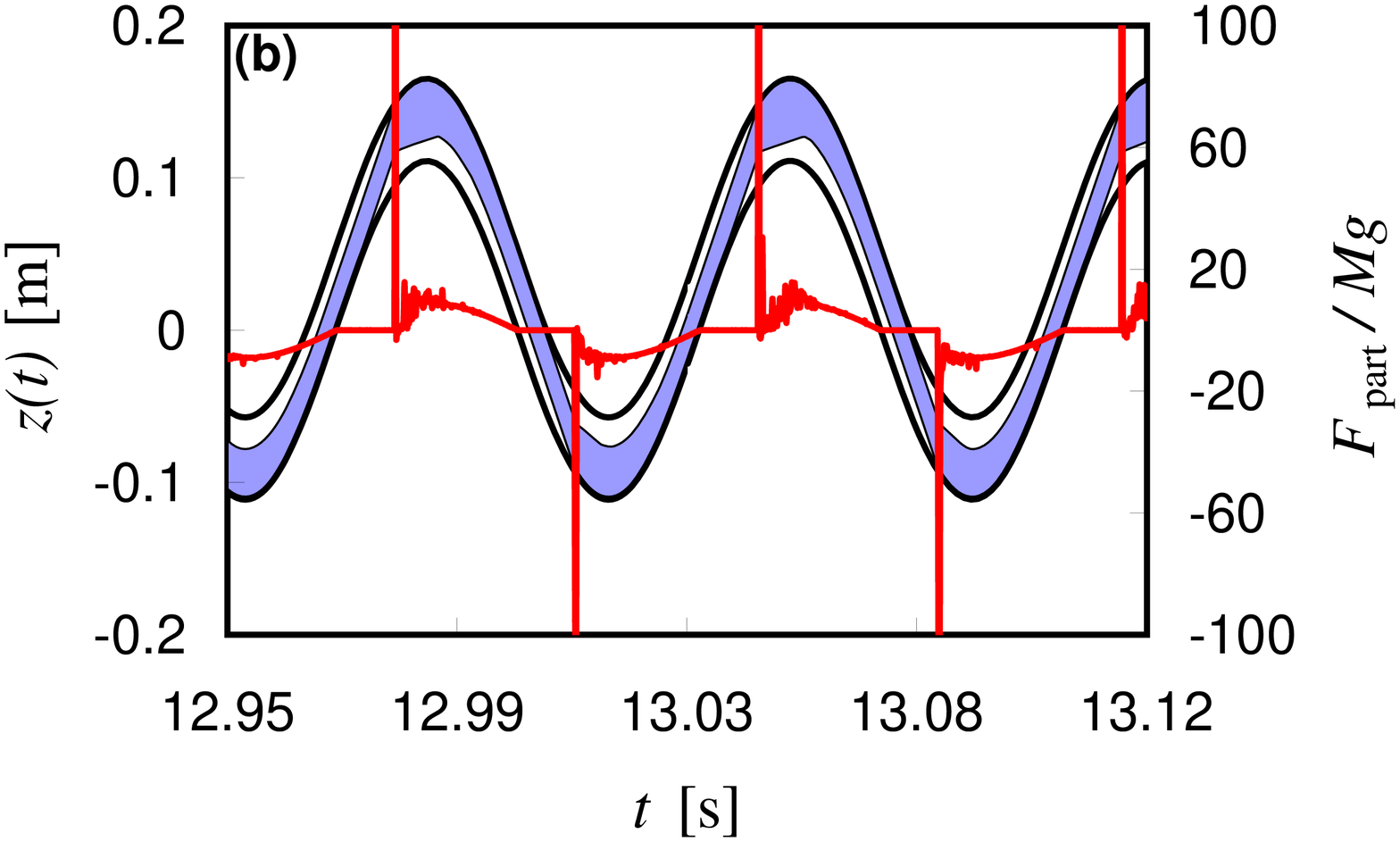}
\includegraphics[width=6cm]{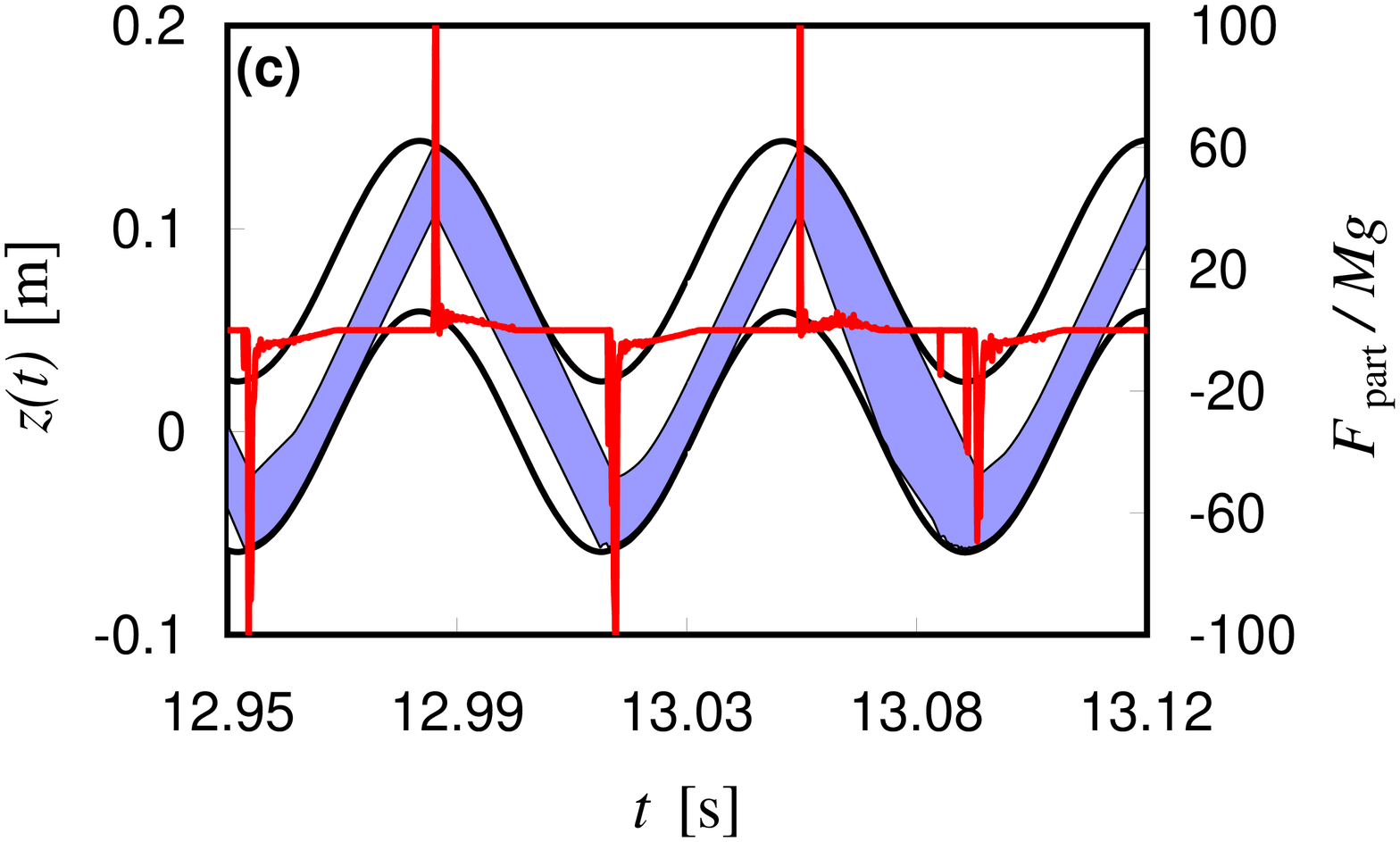}
\includegraphics[width=6cm]{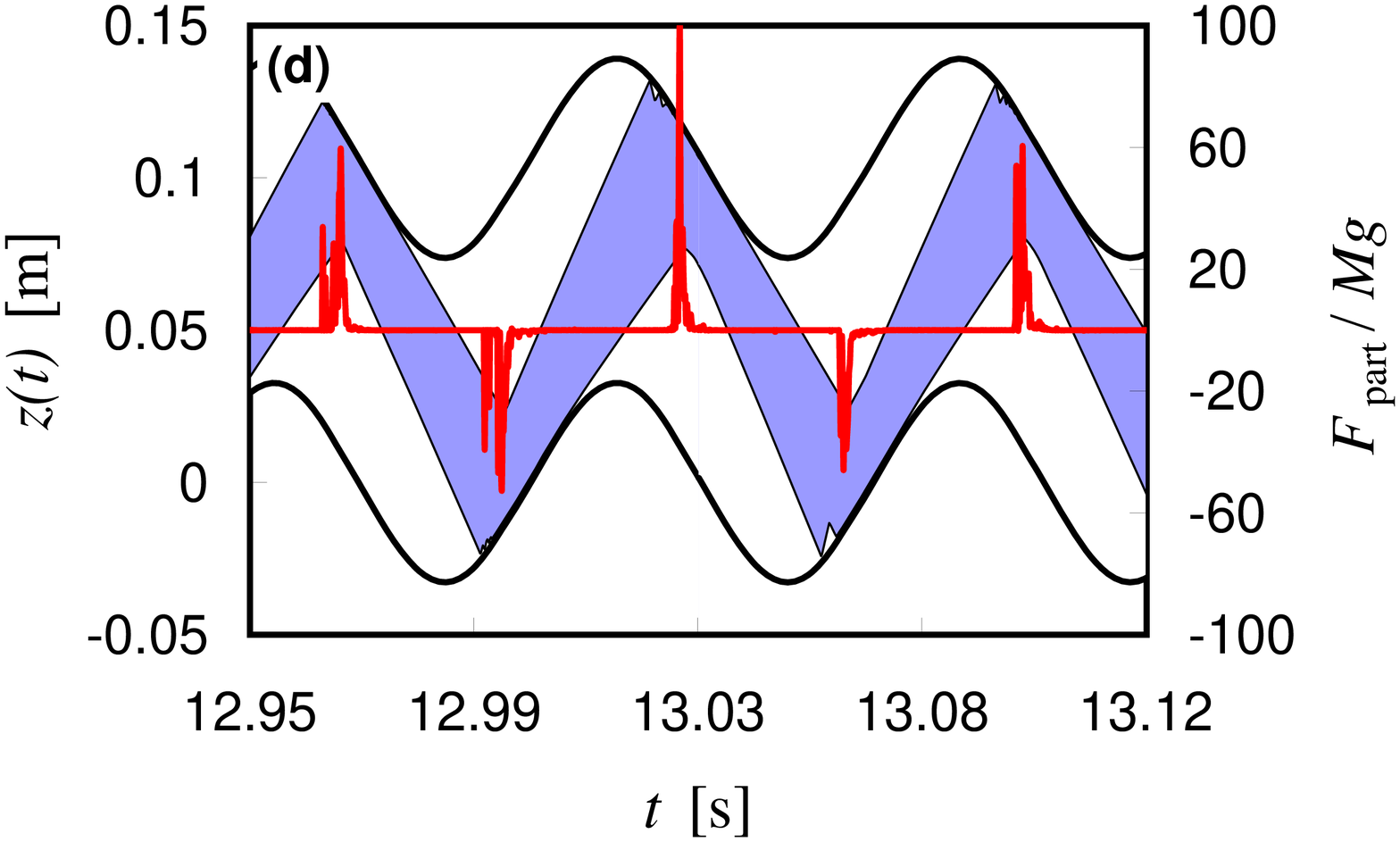}
\includegraphics[width=6cm]{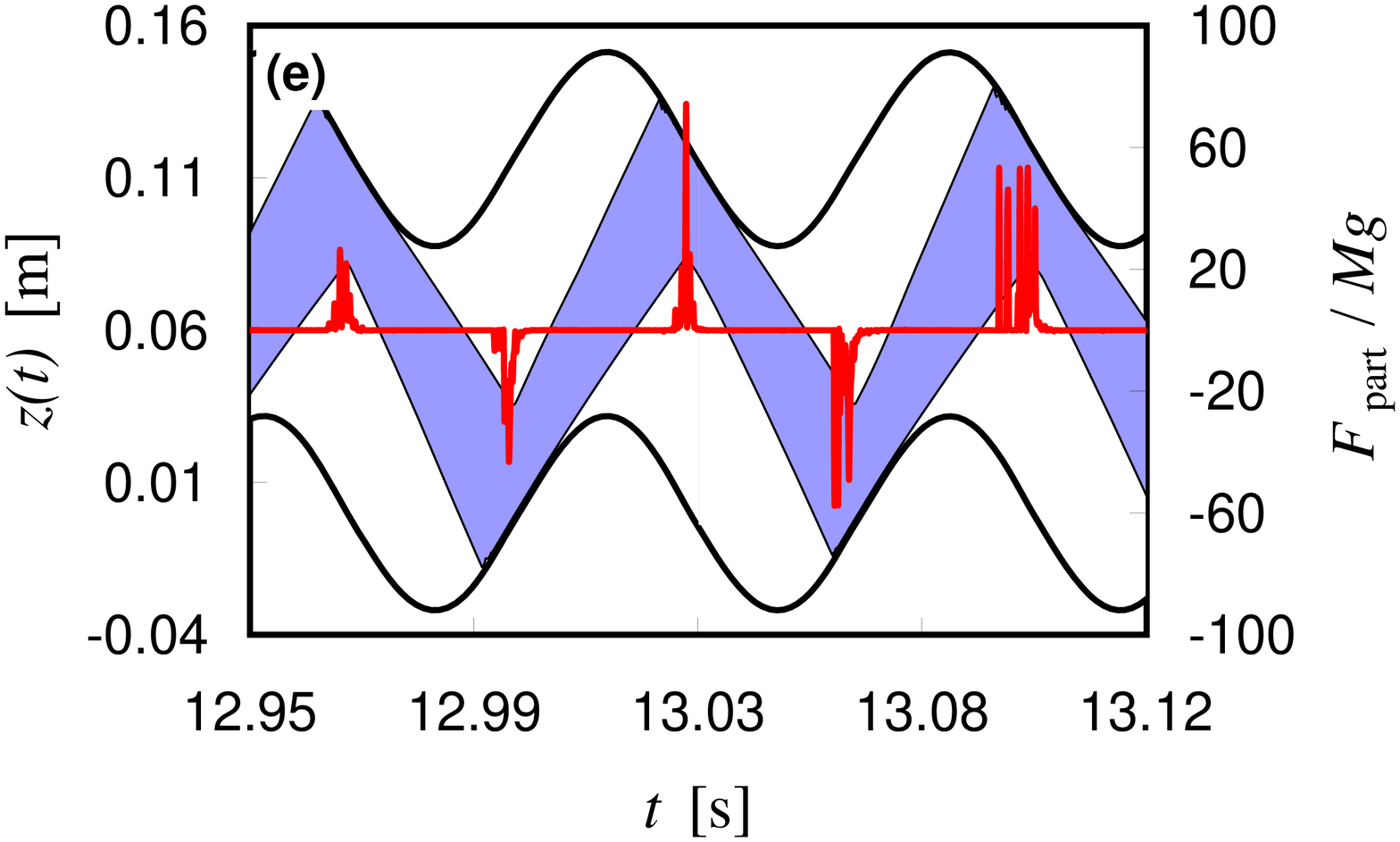}
\includegraphics[width=6cm]{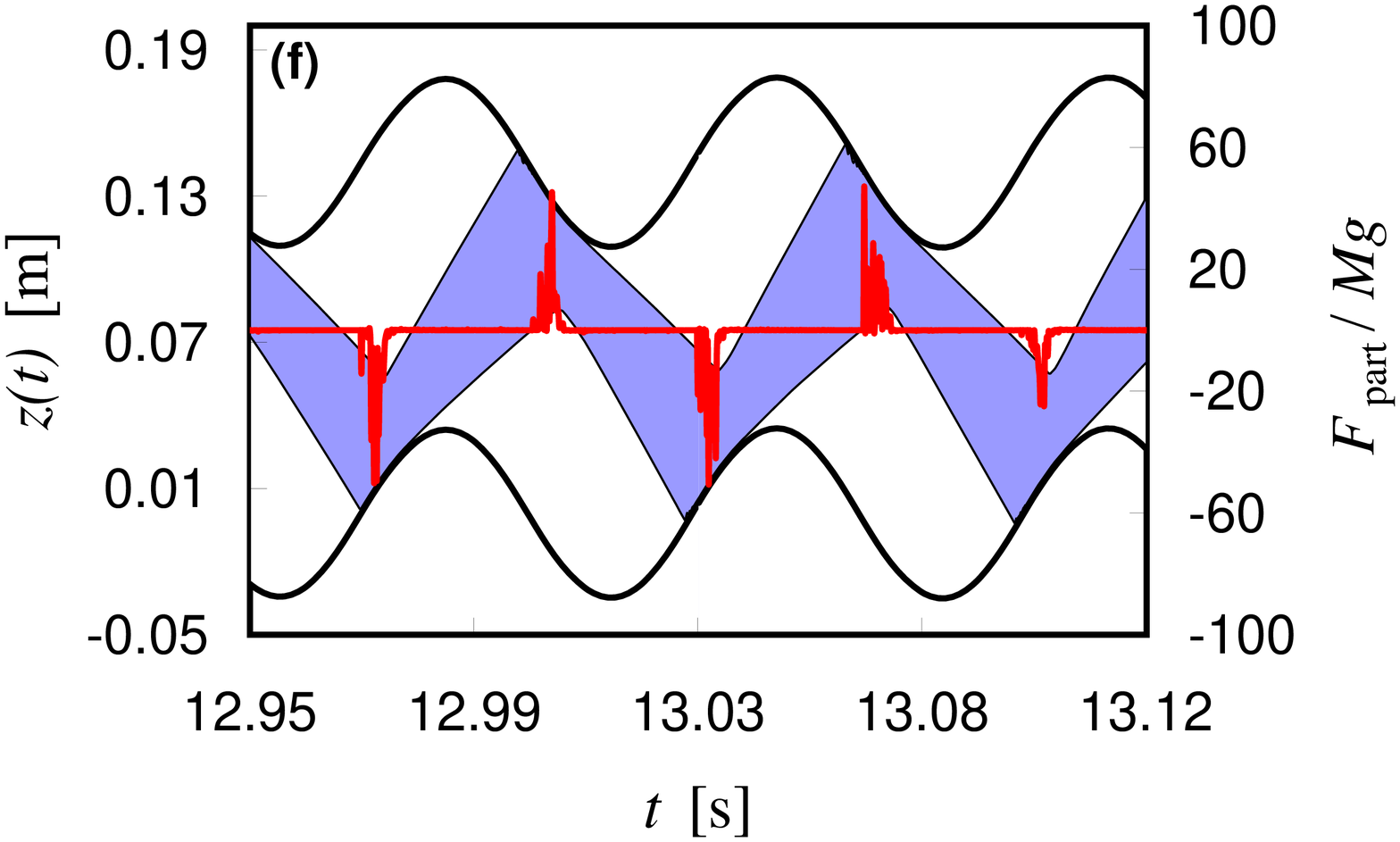}
\includegraphics[width=6cm]{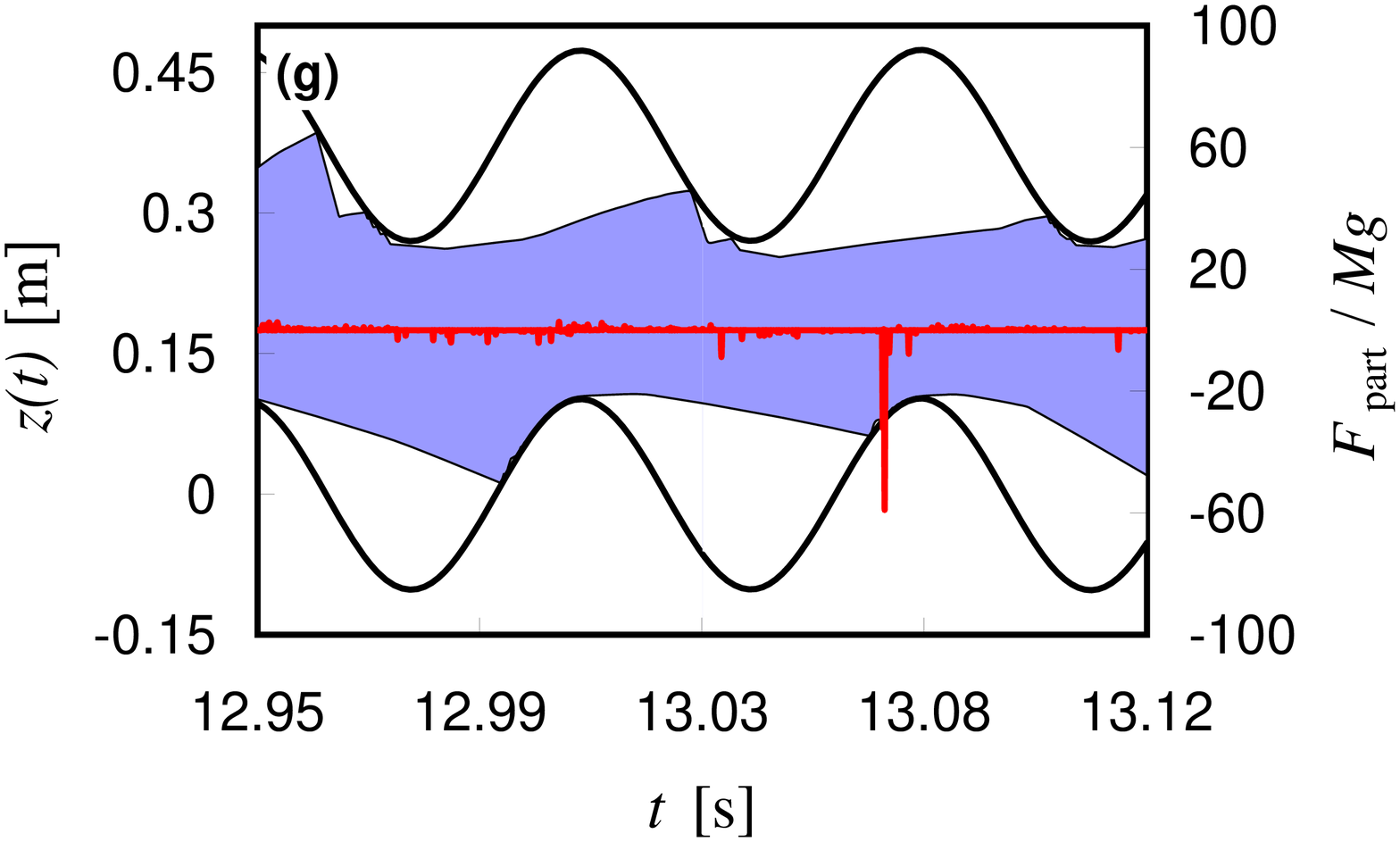}
\includegraphics[width=6cm]{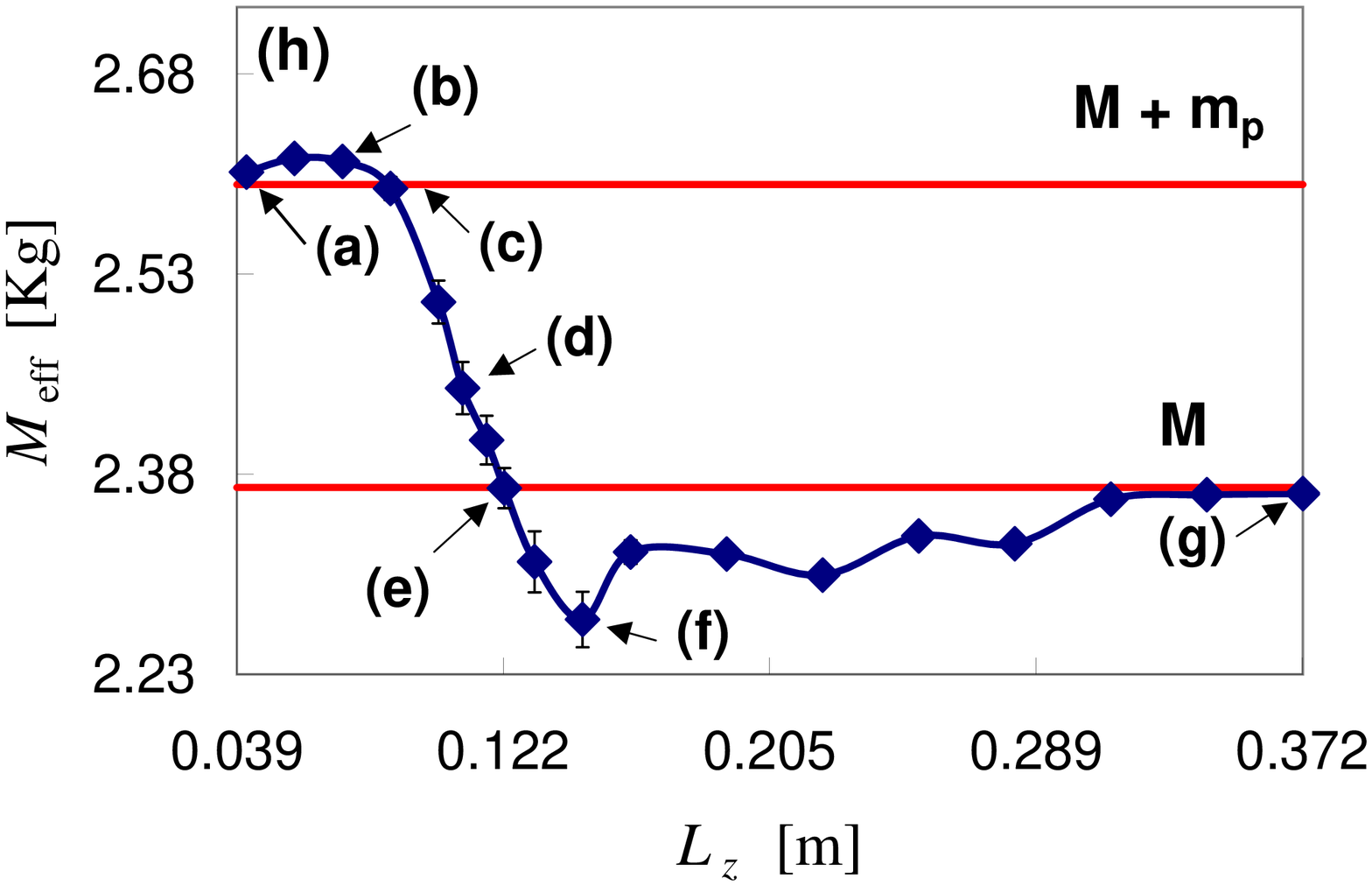}
\end{center}
\caption{Displacement and force of particles against the enclosure for different $L_z$ with $C=7.6$ Nsm$^{-1}$. The solid lines show the position of floor and ceiling of the enclosure and the colored area indicates the limits of the granular bed defined as the position of the uppermost and lowermost particle. \textbf{(a)} $L_z=0.042$ m and $f=14.5$ Hz. \textbf{(b)} $L_z=0.057$ m and $f=14.5$ Hz. \textbf{(c)} $L_z=0.087$ m and $f=14.5$ Hz. \textbf{(d)} $L_z=0.1095$ m and $f=15.0$ Hz. \textbf{(e)} $L_z=0.1225$ m and $f=15.0$ Hz. \textbf{(f)} $L_z=0.147$ m and $f=15.5$ Hz. \textbf{(g)} $L_z=0.372$ m and $f=15.0$ Hz. \textbf{(h)} Same as in Fig. \ref{fig:meff} with arrows indicating the values of $L_z$ corresponding to each panel \textbf{(a)}-\textbf{(g)}.}
\label{fig:fig6}
\end{figure}

In Fig.~\ref{fig:fig6} we plot the trajectory of the floor and ceiling of the enclosure over a few periods of excitation in the steady state regime. We have chosen frequencies close to the resonant frequency for each $L_z$ considered. This is because the effective values of mass and damping obtained by fitting are determined to a large extent by the frequencies with larger amplitude of motion. In Fig.~\ref{fig:fig6}, we have indicated the position of the granular bed inside the enclosure by a band limited by the $z$-coordinates of the uppermost and lowermost particle at any given time. Notice that such representation lets us have an indication of the density of the granular bed and the approximate time of impact with the box top and bottom walls. However, if the granular bed is somewhat dilute, the time of impact of the uppermost or lowermost particle does not coincide with the time at which the most substantial momentum exchange happens between the grains and the enclosure. For this reason we also plot in Fig.~\ref{fig:fig6} the total force exerted by the grains on the enclosure in the $z$-direction (i.e. $F_\mathrm{part}$, see Eq. (\ref{ec2})) which not only provides a more precise assessment of the time of impact but of the intensity of such impacts.

We recall here that the effective mass of an equivalent mass--spring-dashpot model is associated with the response of the PD in phase with the spring force, whereas the effective damping will be associated with the response in phase with the viscous force.

As we mentioned, for very small gaps, the response of the system is similar to the response of an equivalent SDoF system where the total mass of the particles is simply added to the primary mass (i.e., $M_\mathrm{eff}=M+m_\mathrm{p}$). This is due to the fact that the particles are not able to move in such reduced enclosures. As we slightly increase $L_z$, particles behave as a dense pack that travels between the floor and ceiling of the enclosure (see Fig. \ref{fig:fig6}(a)). However, within an oscillation, the granular bed is in full contact with the ceiling or the floor during important portions of the time. During such periods, the granular bed can be considered essentially as an added mass $m_\mathrm{p}$. Notice that after leaving the ceiling (floor) the grains hit the floor (ceiling) before the primary system has reached its maximum displacement. The particles transfer momentum to the enclosure against the direction of the spring force. As a consequence, the effective inertia of the system during the impact is equivalent to a sudden mass increase. This effective mass increase exceeds the loss due to the short periods in which the granular bed is detached from the floor (ceiling) (see the zero force segments in Fig. \ref{fig:fig6}(a)). The overall result is an effective mass $M_\mathrm{eff}$ above $M+m_\mathrm{p}$.

If we further increase $L_z$, the period of detachment within an oscillation increases, but the transfer of momentum at impact also increases leading to an overall increase of $M_\mathrm{eff}$ (see Fig.~\ref{fig:fig6}(b)). Eventually, the transfer of momentum at impact is exactly balanced by the loss of added mass due to the detachment periods. This makes the system render again $M_\mathrm{eff}=M+m_\mathrm{p}$ (see Fig.~\ref{fig:fig6}(c)). Interestingly, this crossover occurs when the granular bed hits the floor (ceiling) at the point of maximum displacement.

There exist a range of values of $L_z$ for which $M<M_\mathrm{eff}<M+m_\mathrm{p}$. In such cases the effective damping is rather high. This is mainly due to the fact that the granular bed hits the enclosure out of phase. In particular, the grains hit the base when both the primary mass is moving upward and the spring force is pulling upward (see Fig.~\ref{fig:fig6}(d)). This results in a strong reduction of the maximum displacement of the system and in an effective added mass with respect to $M$. However, since the periods of detachment from the enclosure are significantly long, the average added mass due to the impacts is smaller than $m_\mathrm{p}$.

If the impacts happens to be always at the time where no spring force is applied to the primary mass, then the transfer of momentum will not result in an effective added mass. This happens when the primary mass pass through the equilibrium point of the spring (i.e., zero displacement). Indeed, we find that there exists a particular value of $L_z$ at which the granular bed hits the floor (ceiling) at this point and the effective mass obtained by fitting corresponds to $M$ (i.e., no added mass, see Fig.~\ref{fig:fig6}(e)). It is important to realize that this is also the value of $L_z$ at which the maximum effective damping is obtained (see Fig.~\ref{fig:ceff} and Ref.\cite{Yang,Friend,Lu}). Therefore, at the optimum $L_z$ where maximum damping is achieved, the effective mass coincides with the primary mass. This means that the addition of particles to create the PD do not affect the resonant frequency of the system if the optimum $L_z$ is chosen, which implies that under such conditions there is no need for compensation of the mass of the particles during design.

As expected, an increase of $L_z$ beyond the optimum damping leads the granular bed to hit the enclosure in phase with the spring force (see Fig.~\ref{fig:fig6}(f)). That is, the grains hit the floor when the spring pulls the system downward. The effect in the apparent inertial response is as if the system suffered a sudden mass decrease. Therefore, the effective mass is smaller than the primary mass (notice that here the granular bed is only in contact with the enclosure during the impacts). 

Much larger gap sizes lead the granular bed to expand significantly. The granular sample enters a gas-like state with only a few particles colliding with the enclosure in each oscillation. This transfers little momentum to the primary mass and the system presents an $M_\mathrm{eff}$ which is close to $M$ (see Fig.~\ref{fig:fig6}(g)).

\section{Conclusions}
\label{concl}
We have studied a PD by means of simulations via a DEM. We have considered the effective mass and effective damping of the entire SDoF system by fitting the FRF to a simple mass--spring--dashpot system. In particular, we study the effect of the height $L_z$ of the enclosure.

We have observed that the effective mass of the system reaches the two limits described in the literature for small and large enclosures. However, those limits are not approached in a monotonous way and clear overshoots appear. For small gap sizes, the system presents effective masses above the direct sum of the primary mass $M$ and the particle mass $m_\mathrm{p}$. For large enclosures, the effective masses fall below $M$.

We have observed that such behavior can be explained by considering both the period of time over which the granular bed is in full contact with the enclosure and the inertial effects due to the grains hitting the floor or ceiling in or out of phase with the spring force.

Interestingly, we found that the value of $L_z$ at which $M_\mathrm{eff}$ crosses $M$ coincides with the optimum damping value described in the literature. Since the optimum $L_z$ should be simple to interpolate from the intersection with the horizontal $M$ level in a plot of $M_\mathrm{eff}$ vs $L_z$, we suggest that such estimation can be a more suitable approach than the search for a maximum in a plot of $C_\mathrm{eff}$ versus $L_z$.

The overshoot effects described are present outside the range of maximal damping performance and might be considered of secondary interest in industrial applications at first sight. However, design constrains may require a PD to work off the optimum damping and in one of the overshoot regions. In particular, for enclosures somewhat taller than the one corresponding to the optimum damping, one achieves effective masses only slightly below the primary mass. This implies that the resonant frequency is almost unaltered upon addition of the particles. Moreover, such values of $L_z$ achieve a remarkable damping (although not maximal) while still presenting a FRF with a shape very similar to the one observed for a simple SDoF mass--spring--dashpot system. This may simplify the prediction of the behavior of the PD under such conditions. On the other hand, if the primary system has more degrees of freedom, the off-design resonant frequencies may fall in either overshoot regime and the side effects should be taken into consideration.

Although we have studied a PD driven in the direction of gravity, similar results are expected if a horizontal setup is considered. It has been shown that near the resonant frequency the response of an impact damper does not depend on the relative direction between the motion of the system and the gravity \cite{Duncan}. Since the effective mass is largely determined by the response near the resonant frequency, the same general trends should be found in horizontally driven PD.

\section*{Acknowledgments}
LAP acknowledges financial support from CONICET (Argentina). 





\bibliographystyle{elsarticle-num}




\end{document}